\documentclass[conference]{IEEEtran}
 \IEEEoverridecommandlockouts
\usepackage[sort,compress]{cite}
\usepackage{amsmath,amssymb,amsfonts}
\usepackage{algorithmic}
\usepackage{multirow}
\usepackage{graphicx}
\usepackage{pifont}
\usepackage{hyperref}
\usepackage{subcaption}
\usepackage{float}
\usepackage{textcomp}
\usepackage{xcolor}
\def\BibTeX{{\rm B\kern-.05em{\sc i\kern-.025em b}\kern-.08em
    T\kern-.1667em\lower.7ex\hbox{E}\kern-.125emX}}

\begin{document}

\title{SoK: Payment Channel Networks
}
\newcommand{\lightning}{\mathsf{LN}}
 \newcommand{\kc}[1]{{[\textcolor{blue}{ kartick: #1}]}}

\author{
    \IEEEauthorblockN{Kartick Kolachala, Mohammed Ababneh, Roopa Vishwanathan}
    \IEEEauthorblockA{
         New Mexico State University, USA \\
         \{kart1712, mababneh, roopav\}@nmsu.edu
    }
}
\pagestyle{plain}
\newcommand{\cmark}{\text{\ding{51}}}
\newcommand{\xmark}{\text{\ding{55}}}

\maketitle

\begin{abstract}
Payment Channel Networks (PCNs) have been proposed as an alternative solution to the scalability, throughput, and cost overhead problems associated with blockchain transactions. By facilitating offchain execution of transactions,  PCNs significantly reduce the burden on the blockchain, leading to faster transaction processing, reduced transaction fees, and enhanced privacy. Despite these advantages, the current state-of-the-art in PCNs presents a variety of challenges that require further exploration.
In this paper, we survey 
several fundamental aspects of PCNs, such as pathfinding and routing, virtual channels, state channels, payment channel hubs, and rebalancing protocols. We aim to provide the reader with a detailed understanding of the various aspects of PCN research, highlighting important advancements. Additionally, we highlight the various unresolved challenges in this area. 
Specifically, this paper seeks to answer the following crucial question: \emph{What are the various interesting and non-trivial challenges in fundamental infrastructure design leading to efficient transaction processing in PCN research that require immediate attention from the academic and research community?} By addressing this question, we aim to identify the most pressing problems and future research directions, and we
hope to inspire researchers and practitioners to tackle these challenges to make PCNs more secure and versatile.

\end{abstract}


\section{Introduction}
\label{sec:intro}

\begin{figure*}[!ht]
    \centering
    \begin{subfigure}[b]{0.30\textwidth}
        \centering
        \includegraphics[width=\textwidth]{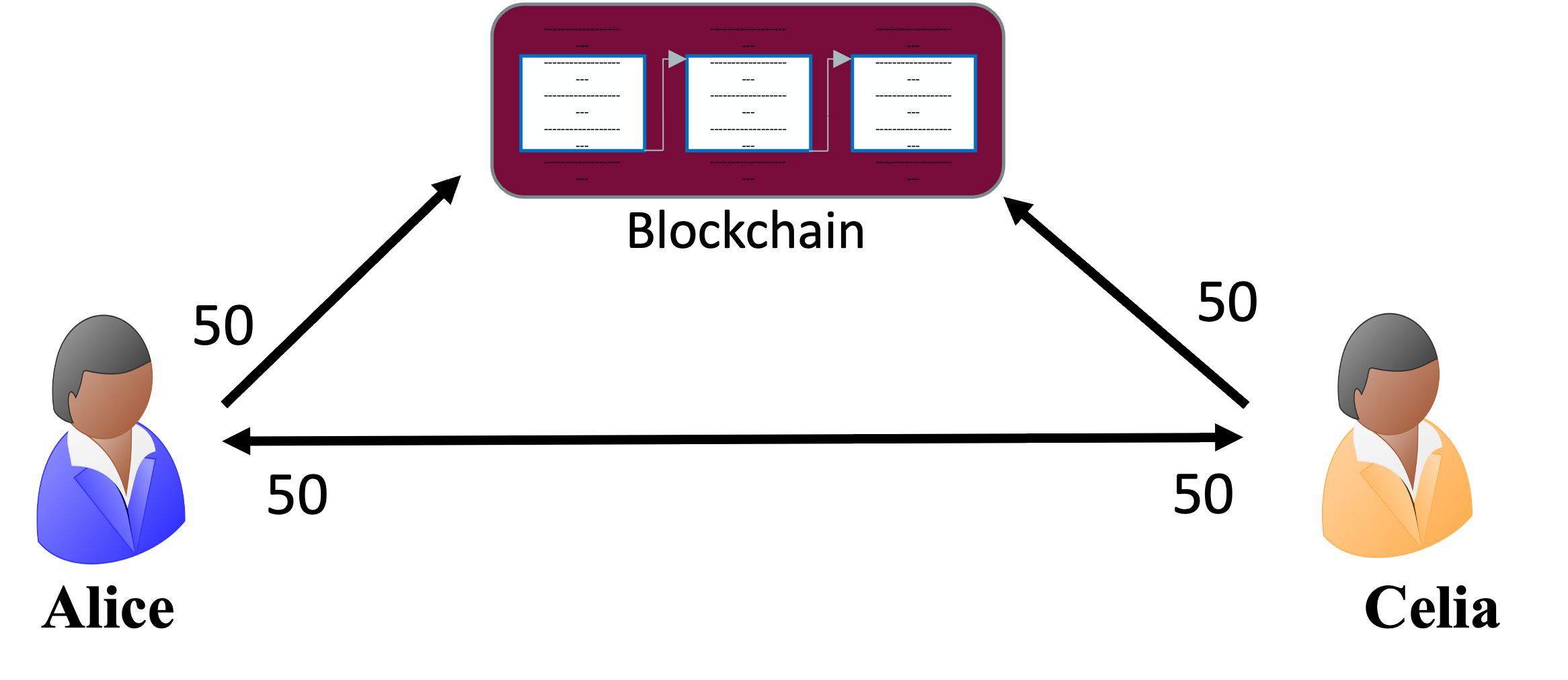}
        \caption{Payment channel opening}
        \label{fig:pcn}
    \end{subfigure}
    \hfill
    \begin{subfigure}[b]{0.30\textwidth}
        \centering
        \includegraphics[width=\textwidth]{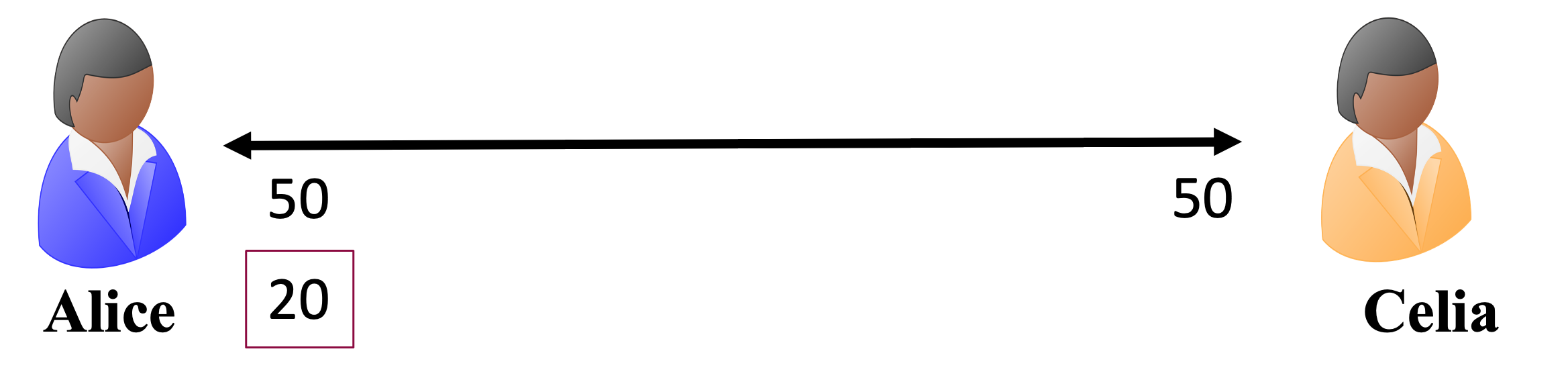}
        \caption{Alice sending 30 coins}
        \label{fig:payment1}
    \end{subfigure}
    \hfill
    \begin{subfigure}[b]{0.30\textwidth}
        \centering
        \includegraphics[width=\textwidth]{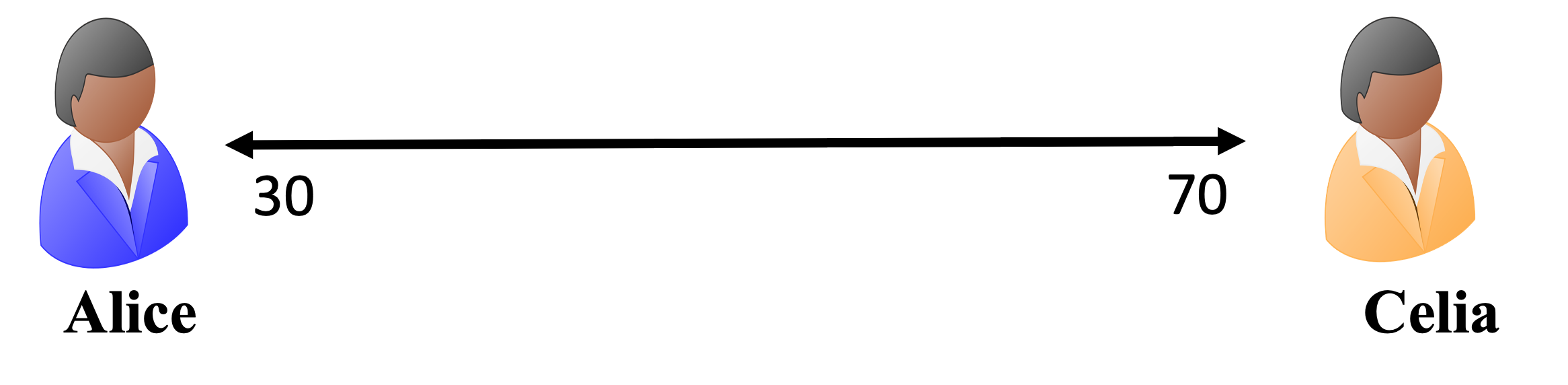}
        \caption{Local balances get updated}
        \label{fig:payment2}
    \end{subfigure}
    \hfill
    \begin{subfigure}[b]{0.30\textwidth}
        \centering
        \includegraphics[width=\textwidth]{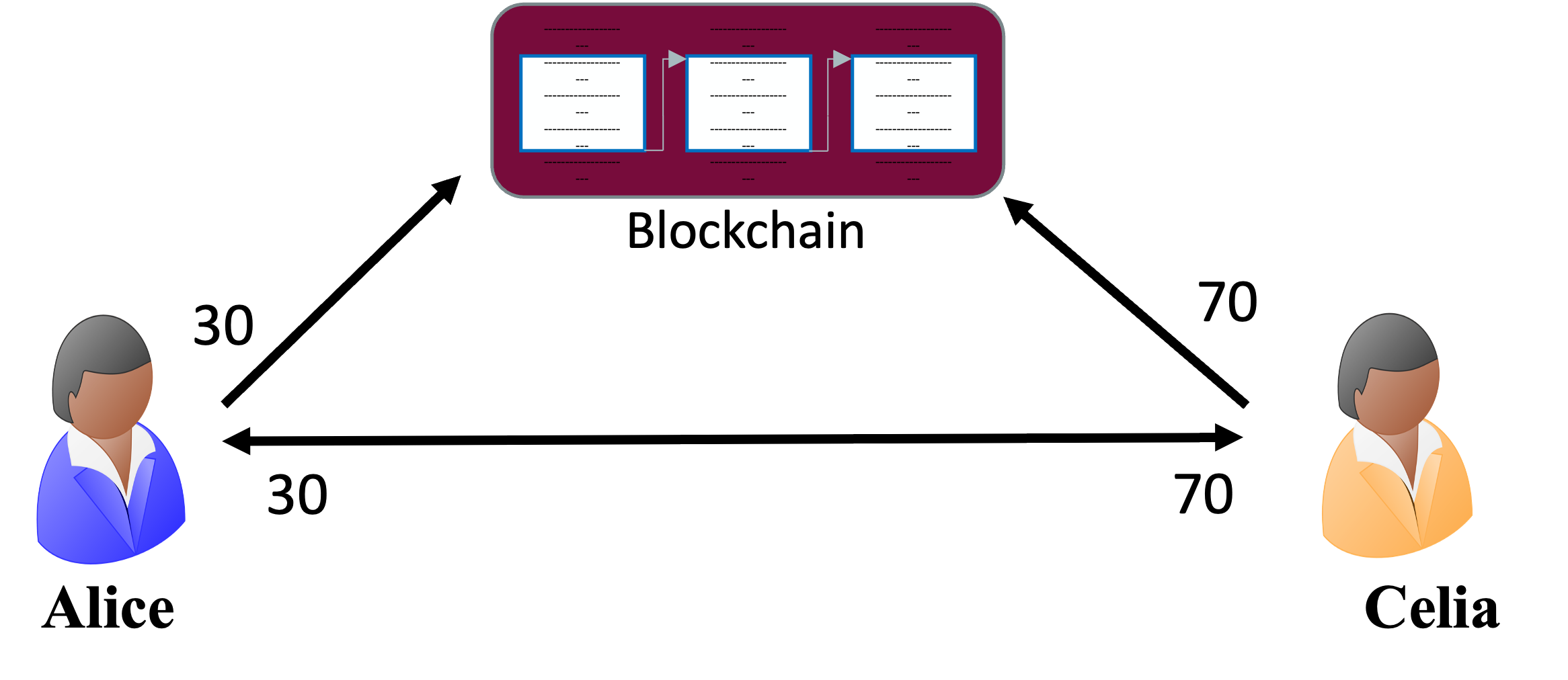}
        \caption{Posting final balances to blockchain}
        \label{fig:close2}
    \end{subfigure}
    \caption{Payment channel operations}
    \label{fig:pc_operations}
\end{figure*}

Cryptocurrencies and cryptocurrency based transactions have become increasing popular. Currently, the total market value of all cryptocurrencies in use has surpassed 2.5 Trillion USD. The cyrptocurrency market is increasing at a rate of $\approx$ 8.00\% every year \cite{forbes}. This rise in popularity can be attributed to the following reasons: 1) cryptocurrency transactions can be carried out without the presence of a trusted entity. Fiat currency transactions on the other hand, require the presence of a trusted financial organization such as a bank. 2) cryptocurrency transactions do not subject the user to any limits on the number and type of transactions. Fiat currency transactions are limited in their amount and number, and depend on several factors such as the currency, geographical location, etc.~\cite{xoomlimit}.\\
Each transaction posted to the Bitcoin blockchain takes around 7 seconds to be validated \cite{btctx, btctx1}. The procedure of validation involves verifying that the transaction posted to the blockchain contains all the required fields and if the signature of the user creating the transaction tuple is valid. Once the validation procedure is successfully completed, the transaction is included in a block that would be mined on the blockchain at some time in the future. The process of mining the block successfully takes $\approx$ 2 hours \cite{btcconfirm} (as of June $13^{th}$ 2024). This delay in the transactions and blocks getting confirmed is termed as the blockchain scalability problem \cite{scale}. In contrast, Visa, a company 
 which globally processes transactions using fiat currencies, processes around 24,000 transactions per second \cite{Visa}. Due to the delay in transaction processing caused by the blockchain scalability problem, blockchain-based transactions cannot process payments instantaneously. \\
As an alternative to processing transactions by posting to the blockchain, payment channels have been proposed. Two parties with the intent of processing payments between them open a payment channel by creating a transaction tuple called the funding transaction. This funding transaction contains the initial deposits from both the parties. These initial deposits are also called as the initial balances of the parties in the payment channel. The sum aggregate of these initial balances is called as the channel capacity. The funding transaction contains the signatures of both the parties involved in the payment channel making it a 2-2 multi signature transaction. This means that the funds in the funding transaction cannot be spent without the signatures of both the parties. This funding transaction is validated and included in a block. Once this block has been successfully mined and confirmed on the blockchain, the payment channel is opened between the two parties. The two parties can now be involved in an unlimited number of transactions with each other as long as the amount of a single transaction does not exceed  their local balances.

An example of a payment channel is given in Figure \ref{fig:pcn}. Two users (also called nodes) Alice and Celia deposit 50 coins each into a 2-2 multi signature transaction. This transaction is posted to the blockchain, upon which it is validated and included in a block. The block is mined and confirmed, at which time a payment channel is said to open between Alice and Celia. The sum aggregate of the individual balances of Alice and Celia, which is the channel capacity, is 100 coins. Alice making a payment of 20 coins to Celia is shown in the Figure \ref{fig:payment1} and the updated balances of Alice and Celia are shown in Figure \ref{fig:payment2}.
After the payment has been made, the channel capacity between Alice and Celia still remains constant at 100 coins. In this manner, Alice and Celia can be involved in an unlimited number of payments between each other. Each payment made in the payment channel in an off-chain manner contains the signatures of both Alice and Celia. When either Alice or Celia decide to close the payment channel, they post their final balances to the blockchain and the payment channel is closed as shown in Figure \ref{fig:close2}.

\begin{figure}[H]
    \centering
    \includegraphics[width=0.5\textwidth]{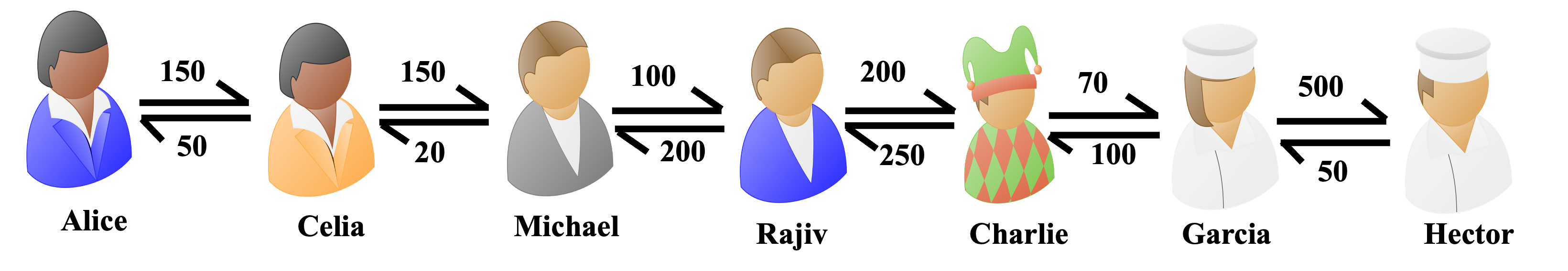}
    \caption{Payment channel network}
    \label{fig:network}
\end{figure}

For each off-chain transaction in the payment channel, both the parties involved in the payment channel create a commitment. This commitment is essentially an agreement for the new balances signed by both the parties. Exchanging of commitments signifies that both the parties have agreed to the change in their respective balances. Each pair of commitments (for each transaction), contains a unique sequence number called the revocation sequence maturity number. For each new transaction made in the payment channel, the sequence number of the prior transaction is invalidated by revocation keys of both the parties. These revocation keys are created by both the parties before opening of the payment channel. If a malicious party in the payment channel broadcasts an older balance to the blockchain, the honest party in the channel has a certain time period during which it can contest this behavior on the blockchain. Before this time period expires, the honest party in the payment channel will broadcast the revocation of this old state signed by both the parties. The broadcasting of this revocation to the blockchain prevents the malicious party from stealing funds of the honest party. \\
The idea of a payment channel that exists between two parties can be extended to a number of nodes, creating a network of payment channels, called a \emph{payment channel network} or PCN. PCNs enable users that are not connected by a direct payment channel to make payments between each other in an off-chain manner. An example PCN is shown in Figure \ref{fig:network}. In the figure, consider Alice who intends to make a payment to Hector, with whom she does not share a payment channel. The na\"ive
 way to process this transaction would be for Alice to open a payment channel with Hector, which involves Alice making an expensive blockchain write for the channel opening. 
Each payment channel opening costs 2.4 USD for blockchain writes \cite{cost1,cost2,cost3}. If Alice intends to send an amount of 1 coin  to Hector, it may not be economical for her to open a direct payment channel. Alice can make use of the PCN and make a payment to Hector by forwarding the payment along the path Alice $\rightarrow$ Celia $\rightarrow$ Michael $\rightarrow$ Rajiv $\rightarrow$ Charlie $\rightarrow$ Garcia $\rightarrow$ Hector. This process of using intermediate nodes in a PCN to forward to the payment to the intended destination is called as routing in payment channel networks.\\
\textbf{Motivation and timeliness of PCN research}: A significant advantage of PCNs is their capability to facilitate micro-payment transactions, with minimum amounts as low as $10^{-7}$ BTC \cite{minlightning}. In contrast, the average transaction cost for a single on-chain transaction on the BTC blockchain is approximately $4.612$ USD as of June 2024 \cite{txcost} \footnote{Opening a payment channel requires two inputs (for funds from both the parties) and one output (the funds are locked in a single 2-2 multi signature address). Whereas making a payment by posting it to the blockchain requires one input (from the party that intends to make the payment) and two outputs (one to the receiver and one to the validator/miner for his fees/reward). This difference makes the transaction cost associated with opening a  payment channel cheaper than making a payment using a blockchain write.}, regardless of the transaction amount. This cost can be avoided by using off-chain PCNs, which incur no additional fees. Additionally, transactions on the BTC blockchain take around 2 hours to be confirmed as of June 13, 2024 \cite{btcconfirm}, whereas PCNs can process transactions instantly. An example of a real-world PCN is the Lightning Network ($\lightning$) on the Bitcoin blockchain \cite{lightning}, which has a 24-hour trading volume of \$63,200 and a market capitalization of \$7 million \cite{lntrade} as of June 2024. These figures reflect $\lightning$'s size and growth.\\ 
In this SoK, we do not survey the various types of attacks in PCNs \cite{avarikioti2024bribe,weintraub2024payout,von2023revelio,khalil2023fakey,mazumdar2022strategic,sguanci,shikhelman2022unjamming,biryukov2022analysis,Riard2021Time,mizrahi2021congestion,weintraub2021structural,perez2020lockdown,harris2020flood,tochner2020route,mazumdar2020griefing,rohrer2020counting,lucongestion,van2020improvements,rohrer2019discharged,romiti2021cross,tikhmirov,vonarx,kappos2021empirical,kumar2023anonymity,zabka2022short,casas,heimbach2024deanonymizing,shikhelman2022unjamming}. The attacks in PCNs usually have overlaps in their strategy and execution, and most of them currently do not have efficient and fully developed mitigation mechanisms proposed. Our conjecture is that, 
their countermeasures might also have design overlaps as and when they are proposed. Hence, we believe attacks in PCNs and their countermeasures require their own taxonomy.
For this SoK, we have covered papers in various areas of PCNs during the time period of 2019-2024 across Tier-1, Tier-2, Tier-3 security conferences in CS, since the norm for security/privacy research and computer science research in general is peer-reviewed conferences.

\textbf{ Contributions}: \\
1) We qualitatively compare the recent work in several aspects of building PCNs, viz. pathfinding and routing, virtual channels, state channels, payment channel hubs, and rebalancing using several relevant properties (metrics) along with providing a reasoning why these metrics have been chosen for comparison.\\
2) We point out the open problems in all the areas that we survey and we also discuss why solving those problems is a hard research challenge.

\textbf{Outline}: 
In Section \ref{sec:routing}, we start  with describing the concept of pathfinding and routing in PCNs and qualitatively compare work published in that area. In Section \ref{sec:tumbler}, we describe virtual channels which have been proposed to address issues with multi-hop routing in PCNs, and compare work in this area. 
In Section \ref{sec:state}, we describe and compare state channel protocols, which are a more generalized adaptation of virtual channels and can facilitate the execution of arbitrary applications between nodes in the PCN (not just payments). 
In Section \ref{sec:tumbler}, we cover payment channel hubs, which are similar to virtual channels and state channels, but facilitate a different use case for payments in PCNs. 
All of the aforementioned PCN mechanisms consume one common resource: the local balance of a node in a PCN. In Section \ref{sec:rebal}, we discuss rebalancing, which addresses the important function of replenishing channel funds in the PCN. In Section \ref{sec:gaps}, we present the reader with the current research gaps in all of the aforementioned areas and also describe why bridging those gaps is hard. In Section \ref{sec:conc}, we conclude the paper.

\textbf{Prior work}: Prior works by Khojasteh \emph{et al.} \cite{khojasteh2021survey} and Erdin \emph{et al.} \cite{erdin2102evaluation} survey the work done only in the area of pathfinding and routing protocols and their privacy aspects in PCNs.
Whereas, in this paper we cover the entire spectrum of PCN research: 
rebalancing, virtual channels, state channels, pathfinding and routing, and tumblers. Apart from this~\cite{khojasteh2021survey,erdin2102evaluation}
do not provide any information about the open problems in PCNs, which we do in our work. The SoK by Gudgeon \emph{et al.} \cite{gudgeon2020sok}, surveys several layer-2 protocols, whereas, we focus exclusively on PCNs. Furthermore~\cite{gudgeon2020sok} was published in 2020 and does not cover most of the recent work published in PCNs.

\begin{table*}[h!]
\centering
\caption{Comparison of Pathfinding and Routing Protocols in PCNs}
\label{tbl:routing}
\begin{tabular}{|p{2.50cm}|p{1.50cm}|p{0.75cm}|p{1.30cm} |p{1.00cm}|p{1.75cm}|p{1.25cm}|p{1.00cm}|p{1.55cm}|p{1.00cm}|}
\hline
 Protocols                                                                   & Concurrency  & Privacy      & \begin{tabular}[c]{@{}l@{}}Topology\\  privacy\end{tabular} & \begin{tabular}[c]{@{}l@{}}Avoids \\ source \\ routing\end{tabular} & Decentralized & Atomicity    & \begin{tabular}[c]{@{}l@{}}Disjoint \\ graphs\end{tabular} & Fees & Year                   \\ \hline\hline
 SilentWhispers \cite{Malavolta2017SilentWhispersES}              & $\xmark$     & $\checkmark$ & $\checkmark$                                                & $\checkmark$                                                        & $\checkmark$  & $\xmark$     & $\xmark$                                                   & $\xmark$  & 2017             \\ \hline
 SpeedyMurmurs \cite{speedy}                                      & $\checkmark$ & $\checkmark$ & $\checkmark$                                                & $\checkmark$                                                        & $\checkmark$  & $\xmark$     & $\xmark$                                                   & $\xmark$ & 2018              \\ \hline
 Coinexpress \cite{coinexpress}                                   & $\checkmark$ & $\xmark$     & $\checkmark$                                                & $\checkmark$                                                        & $\checkmark$  & $\checkmark$ & $\xmark$                                                   & $\xmark$ & 2018               \\ \hline
 Blanc \cite{blanc}                                                & $\checkmark$ & $\checkmark$ & $\checkmark$                                                & $\checkmark$                                                        & $\checkmark$  & $\checkmark$ & $\xmark$                                                   & $\xmark$ & 2019               \\ \hline
 Robustpay \cite{robustpay}                                        & $\xmark$     & $\xmark$     & $\xmark$                                                    & $\xmark$                                                            & $\xmark$      & $\checkmark$ & $\xmark$                                                   & $\checkmark$ (flat) & 2019   \\ \hline
 Flash \cite{flash}                                                & $\xmark$     & $\xmark$     & $\xmark$                                                    & $\xmark$                                                            & $\xmark$      & $\checkmark$ & $\xmark$                                                   & $\xmark$ & 2019              \\ \hline
 Cheapay \cite{cheapay}                                                 & $\xmark$     & $\checkmark$ & $\checkmark$                                                & $\checkmark$                                                        & $\checkmark$  & $\xmark$     & $\xmark$                                                   & $\checkmark$ (flat) & 2019  \\ \hline
 Eckey \emph{et al} \cite{Eckey2020SplittingPL} & $\xmark$     & $\xmark$     & $\checkmark$                                                & $\checkmark$                                                        & $\checkmark$  & $\checkmark$ & $\xmark$                                                   & $\xmark$ & 2020              \\ \hline
 FSTR \cite{FSTR}                                                 & $\xmark$     & $\xmark$     & $\xmark$                                                    & $\xmark$                                                            & $\xmark$      & $\xmark$     & $\xmark$                                                   & $\xmark$ & 2020              \\ \hline

Spider \cite{spider}                                          & $\xmark$     & $\xmark$     & $\xmark$                                                    & $\xmark$                                                            & $\xmark$      & $\xmark$     & $\xmark$                                                   & $\xmark$ & 2020               \\ \hline
Vein \cite{vein}                                                  & $\xmark$     & $\xmark$     & $\xmark$                                                    & $\xmark$                                                            & $\xmark$      & $\xmark$     & $\xmark$                                                   & $\checkmark$ (dynamic) & 2021 \\ \hline
Kadry \emph{et al.} \cite{machine}        & $\xmark$     & $\xmark$     & $\xmark$                                                    & $\xmark$                                                            & $\xmark$      & $\xmark$     & $\xmark$                                                   & $\xmark$ & 2021              \\ \hline

Webflow \cite{webflow}                                            & $\xmark$     & $\checkmark$ & $\checkmark$                                                & $\checkmark$                                                        & $\checkmark$  & $\xmark$     & $\xmark$                                                   & $\xmark$ & 2021              \\ \hline
Robustpay+ \cite{Zhang2021RobustPayRP}                            & $\xmark$     & $\xmark$     & $\xmark$                                                    & $\xmark$                                                            & $\xmark$      & $\checkmark$ & $\xmark$                                                   & $\checkmark$ (flat) & 2021   \\ \hline

MPCN-RP \cite{mpcnrp}                                             & $\xmark$     & $\xmark$     & $\xmark$                                                    & $\xmark$                                                            & $\xmark$      & $\checkmark$ & $\xmark$                                                   & $\checkmark$ (flat) & 2022    \\ \hline

Auto tune \cite{hong2023auto}                                     & $\xmark$     & $\xmark$     & $\xmark$                                                    & $\xmark$                                                            & $\xmark$      & $\xmark$     & $\xmark$                                                   & $\checkmark$ (flat) & 2023    \\ \hline
Yang \emph{et al.} \cite{optimalhub}                  & $\checkmark$ & $\checkmark$ & $\checkmark$                                                & $\checkmark$                                                        & $\xmark$      & $\xmark$     & $\xmark$                                                   & $\xmark$ & 2023               \\ \hline

RACED \cite{kolachala2023raced}                                        & $\checkmark$ & $\checkmark$ & $\checkmark$                                                & $\checkmark$                                                        & $\checkmark$  & $\checkmark$ & $\checkmark$                                               & $\xmark$  & 2024             \\ \hline

Auroch \cite{ababneh2024auroch}                                        & $\checkmark$ & $\checkmark$ & $\checkmark$                                                    & $\checkmark$                                                        & $\checkmark$  & $\checkmark$ & $\xmark$                                                   & $\checkmark$ (dynamic) & 2024          \\ \hline

SPRITE \cite{panwar2024sprite}                                         & $\checkmark$ & $\checkmark$ & $\checkmark$                                                & $\checkmark$                                                        & $\checkmark$  & $\checkmark$ & $\xmark$                                                   & $\xmark$ & 2024               \\ \hline
\end{tabular}
\end{table*}

\section{Pathfinding and Routing}
\label{sec:routing}


\textbf{Motivation}: One of the areas in PCNs that has garnered significant attention from the academic community is pathfinding and routing. Pathfinding is defined as the process of finding a path comprising several nodes from a sender to a receiver in a PCN along which a payment 
can potentially be forwarded, 
and routing is the process of actually forwarding the payment along the found path. Intuitively, it may seem that well-known pathfinding and routing protocols from the wired and wireless networks domain can be easily applied to PCNs. Unfortunately, there are several problems with this: 
1) Traditional networks focus on the transfer of data, PCNs on the other hand, transfer money in a decentralized manner. 2) Data transfer in traditional peer-peer networks does not alter the bandwidth, whereas money transfer in PCNs alters the monetary state of the nodes involved. 3)  Cost in traditional networks is measured in terms of latency, whereas in PCNs, it involves routing fees, leading to greedy behavior among users and makes PCNs vulnerable to various attacks \cite{gudgeon2020sok}.

The properties on the columns in Table \ref{tbl:routing} represent the fundamental principles of fiat currency transactions and on-chain payment mechanisms, which we want reflected in off-chain payments. These properties are generally agreed upon in the literature by several works such as \cite{kolachala2023raced,ababneh2024auroch,panwar2024sprite,blanc,speedy,Malavolta2017SilentWhispersES} as common evaluation metrics for pathfinding and routing protocols in PCNs. Fulfilling these properties while providing efficient pathfinding and routing is a non-trivial challenge, and necessitates the design of new pathfinding and routing protocols. 
Several elegant pathfinding and routing protocols have been proposed in the literature. 
In Table \ref{tbl:routing}, we present a qualitative comparison of these routing protocols with respect to the properties they achieve. In this paper across all sections, if any prior work has identified a property as ideal or has identified a gap in research, we give an appropriate citation(s). If there is no citation provided, it indicates that the corresponding property/research gap has been identified by us.

\textbf{Ideal properties}: 
1) \textbf{Concurrency}: Concurrency is achieved when a pathfinding and a routing protocol enables the nodes to forward more than one payment simultaneously \cite{malavolta2017concurrency}.  
\emph{Importance}: At a given instant of time, many users will be using the PCN to make offchain payments. Hence it is important for a routing protocol to support concurrency.
2) \textbf{Privacy}: Privacy is maintained when a node's real identity is known only to its immediate neighbors and not to the entire network. 
\emph{Importance}: Information of a node such as its identity, local balance, connections with other nodes in the network and its transaction history are private and should not be known to anyone else.
3) \textbf{Topology privacy}: Topology privacy is preserved when no single node has knowledge of the entire network topology. 
\emph{Importance}: If topology privacy is not preserved, it violates the privacy of every node in the network. Making network topology public can potentially lead an adversary to reconstruct transaction paths, which in turn can lead to an adversary selectively targeting a certain set of nodes.
4) \textbf{Avoids source routing}: Source routing is avoided when the sender does not determine the path to the receiver. 
\emph{Importance}: If a sender determines the complete path to the receiver, it means that he has access to the entire network topology. PCNs are highly dense and dynamic in nature. It is practically infeasible for a node to maintain an updated network topology all the time. 
5) \textbf{ Decentralization}: Decentralization is achieved when there are no centralized, trusted entities responsible for constructing paths for senders. 
\emph{Importance}: Cryptocurrency payments made using the blockchain are by nature decentralized, hence routing protocols which facilitate offchain cryptocurrency payments should also be decentralized.
6) \textbf{Atomicity}: Atomicity is ensured when the payment is routed all the way from the sender to the receiver, or the payment is not routed at all. 
\emph{Importance}: Atomicity is important since it ensures that honest people do not lose their funds because of malicious behavior by other parties in the system.  
7) \textbf{Disjoint graphs}: A pathfinding and routing protocol is considered applicable to disjoint graphs if it functions even when the network graph consists of islands. 
\emph{Importance}: PCNs often comprise of islands which only have a couple of nodes. 
These islands are often disconnected from other dense parts of the PCN. 
A routing protocol should be able to facilitate transactions between any pair of nodes irrespective of their location.
8) \textbf{Fees}: Routing fees is the amount a node charges for forwarding the payment to the next node along a path from the sender to receiver. This fees can be charged in two ways. Flat/fixed fees means that the fees charged for routing payments remains the same irrespective of the transaction amount being routed. If the fees charged by a node varies according to the transaction amount, it is referred to as dynamic fees, typically a percentage of the amount.  
\emph{Importance}: Every node along a payment path aids the sender in transaction processing by forwarding the payment to the next node along the path. The nodes need to be paid in the form of routing fees for their service.

As illustrated in Table \ref{tbl:routing}, routing protocols for PCNs have evolved significantly over the years. The two most significant advancements are taking routing fees into consideration and providing support for privacy. For instance, $\lightning$ provides users with two sets of keys, a long-term keypair and an alias (a temporary identity) helping to conceal their identities and ensure privacy. Despite these developments, two overarching research problems remain that require attention. We discuss them in detail in Section \ref{sec:gaps}.


\section{Virtual channels}
\label{sec:virtual}
\begin{figure*}[h!]
    \centering
    \begin{subfigure}[t]{0.45\textwidth}
        \centering
        \includegraphics[width=\textwidth]{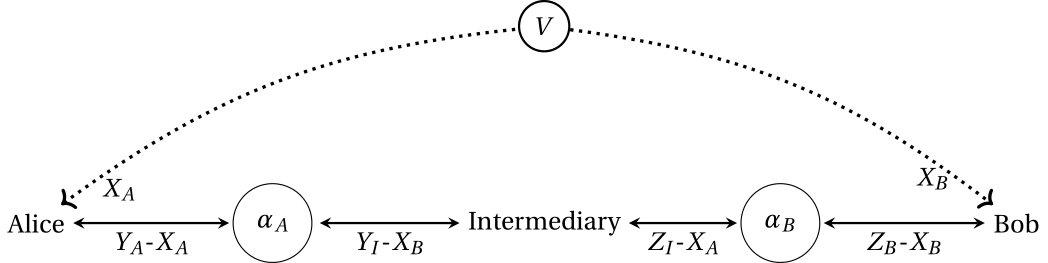}
        \caption{Virtual channel example \cite{dziembowski2019perun,AML}}
        \label{fig:perun}
    \end{subfigure}
    \hfill
     \begin{subfigure}[t]{0.45\textwidth}
        \centering
        \includegraphics[width=\textwidth]{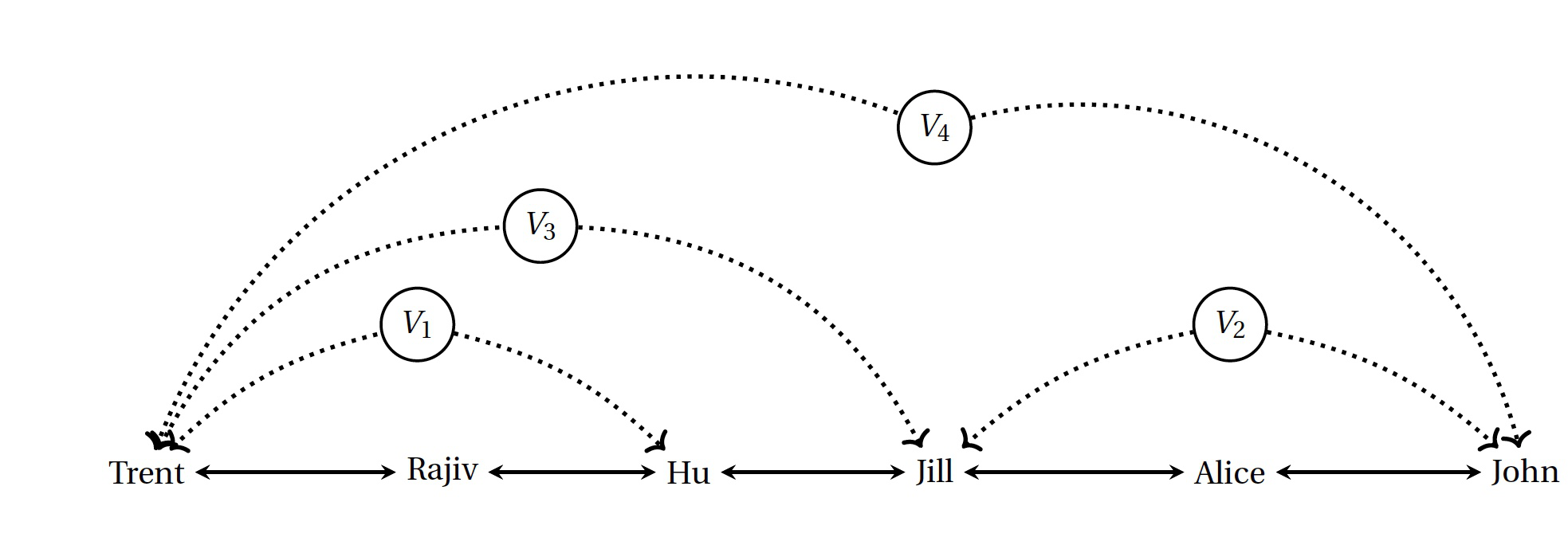}
        \caption{Recursive virtual channel}
        \label{fig:recursive}
    \end{subfigure}
    \hfill
    \begin{subfigure}[t]{0.45\textwidth}
        \centering
        \includegraphics[width=\textwidth]{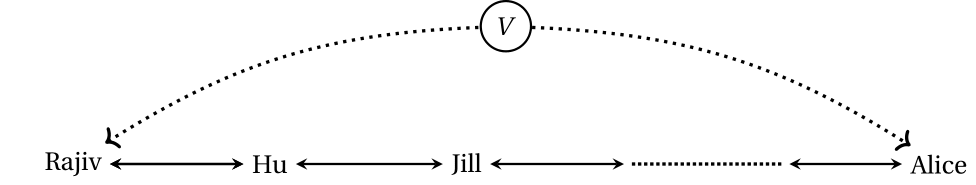}
        \caption{Multi-hop virtual channel (the dots between Jill and Alice depict several intermediate nodes)}
        \label{fig:multihop}
    \end{subfigure}
   
    \caption{Different types of virtual channels}
    \label{fig:combined}
\end{figure*}
\begin{table*}[h!]
\centering
\caption{Comparison of Virtual Channel Protocols}
\label{tbl:virtual}
\begin{tabular}{|p{1.70cm}|p{1.00cm}|p{1.00cm}|p{2.00cm}|p{1.00cm}|p{1.40cm}|p{1.00cm}|p{1.00cm}|p{1.00cm}|}
\hline
\multicolumn{1}{|c|}{Protocol}                                                          & BC          & Validity  & Fee                                                                     & Privacy      & \begin{tabular}[c]{@{}c@{}}Off-chain \\ dispute \\ resolution\end{tabular}    & Recursive    & Multihop & Year \\ \hline
Generalized state channels \cite{dziembowski2018general}               & TC          & Limited   & $\xmark$                                                                & $\xmark$     & $\xmark$                                                                 & $\checkmark$ & $\xmark$ & 2018 \\ \hline
Eckey \emph{et al.} \cite{dziembowski2019multi}       & TC          & Limited   & $\xmark$                                                                & $\xmark$     & $\xmark$                                                                & $\checkmark$ & $\xmark$ & 2019 \\ \hline
Perun \cite{dziembowski2019perun}                                 & TC          & Limited   & $\xmark$                                                                & $\checkmark$     & $\xmark$                                                                     & $\xmark$     & $\xmark$ & 2019\\ \hline
Jourenko \emph{et al.} \cite{jourenko2020lightweight} & UTXO        & Limited   & $\xmark$                                                                & $\xmark$     & $\xmark$                                                                 & $\xmark$ & $\xmark$ & 2020 \\ \hline
Aumayr \emph{et al.} \cite{aumayr2021bitcoin}         & UTXO        & Limited   & $\checkmark$ (fixed)     & $\xmark$     & $\xmark$                                                                     & $\xmark$     & $\xmark$ & 2021 \\ \hline
Elmo \cite{kiayias2021elmo}                                            & UTXO        & Unlimited & $\xmark$                                                                & $\xmark$     & $\xmark$                                                                     & $\checkmark$ & $\xmark$ & 2021 \\ \hline
Donner \cite{aumayr2023breaking}                                       & UTXO        & Unlimited & $\checkmark$ (time based) & $\checkmark$ & $\xmark$                                                                 & $\checkmark$ & $\xmark$ & 2023 \\ \hline
Jia \emph{et al.} \cite{jia2023cross}                 & UTXO, TC & Limited   & $\checkmark$ (fixed)                                                                & $\xmark$     & $\xmark$                                                             & $\xmark$     & $\xmark$ & 2023 \\ \hline

\end{tabular}
\end{table*}
\textbf{Motivation}:
Transactions in PCNs are routed from the sender to the receiver using a path of intermediate nodes.
Current pathfinding and routing mechanisms require the nodes along the payment path to be available for a transaction to be processed. However, nodes can sometimes choose to go offline or there can be network/service disconnections causing transaction failures. Furthermore, each node along a payment path charges its own fees for forwarding the payment, which is paid by the sender and increases with the path length, hence the time taken to route a payment grows linearly in the path length. 
Virtual channels, which are built on top of existing payment channels, solve these problems. Initial constructions of virtual channels facilitated payments between a pair of nodes using a single intermediate node \cite{dziembowski2019perun}.
The intermediate node needs to have individual payment channels open with the other two nodes. The intermediary and the pair of nodes lock coins with each other in their respective payment channels and a virtual channel is established. Upon establishment of the virtual channel, the pair of nodes can be involved in a unlimited number of payments. These payments can be processed without the intermediate node being online. It might be better to use virtual channels 
from a routing fees perspective, since unlike routing protocols, nodes in virtual channels do not charge a routing fee for every transaction.

Alice, Bob, and an intermediary establish a virtual channel as shown in Figure \ref{fig:perun}. Alice locks $Y_{A}$ coins and the intermediary locks $Y_{I}$ coins in the payment channel $\alpha_{A}$. Similarly, Bob locks $Z_B$ coins and the intermediary locks $Z_I$ coins in their channel $\alpha_{B}$. The virtual channel $V$ is created once Alice locks $X_{A}$ coins from her balance in $\alpha_{A}$ and Bob locks $X_{B}$ coins from his balance in $\alpha_{B}$. Now Alice and Bob can process payments without the intermediary's online presence.

The idea of a virtual channel between a pair of nodes involving a single intermediary has been extended to establishing a virtual channel recursively over several hops involving several intermediaries, leading to the construction of a recursive virtual channel \cite{jourenko2020lightweight,kiayias2021elmo,aumayr2023breaking}. An example of a recursive channel is shown in Figure \ref{fig:recursive}. A recursive virtual channel enables transaction processing between any pair of nodes, by recursively establishing virtual channels between several intermediaries, as opposed to~\cite{dziembowski2019perun}, which facilitates virtual channels only between a pair of nodes connected directly to an intermediary.


A multihop virtual channel is constructed by establishing a single virtual channel  between a pair of users over a path comprising of several intermediate nodes. An example of multihop virtual channel is shown in Figure~\ref{fig:multihop}. A multihop virtual channel is an improvement over recursive virtual channels. In a multihop virtual channel, a single virtual channel can be established between any pair of nodes which are separated by several intermediaries.

Virtual channels should not add an unnecessary burden to users, and should mirror the operations of payment channels as closely as possible, with the added benefit of no on-chain transactions at all, while maintaining comparable security/privacy properties. We now give the properties desired from an ideal virtual channel, and compare the works in this area on the extent to which they achieve these properties. 

\emph{Blockchain terminology}: In the rest of the paper BC denotes a blockchain.
TC is a  blockchain that supports a Turing complete programming language, such as Ethereum, and UTXO is a blockchain that supports the UTXO-based scripting mechanism such as Bitcoin. In Table \ref{tbl:virtual}, we present a qualitative comparison of virtual channel protocols.



\textbf{Ideal properties}: 
1) \textbf{Validity}: This determines the validity of the virtual channel. A limited validity means that the virtual channel is valid for a predetermined time period (which is decided by nodes involved in the virtual channel). Unlimited validity means that the virtual channel can stay open until the nodes involved decide to initiate closing \cite{dziembowski2019multi}. 
\emph{Importance}: Unlimited validity provides better and efficient transaction processing between users who intend to have frequent payments made between each other. 
2) \textbf{Fee}: This metric determines if the virtual channel takes into account the fees charged by the intermediate node(s) involved in the channel's establishment. The fixed fee model implies that a predetermined, fixed fee is paid to the intermediate node(s) which is agreed upon by all the nodes in the virtual channel before the channel establishment.
The time-based fee model implies that the fee paid to the intermediary depends on the time for which the virtual channel stays open \cite{dziembowski2019multi}.
\emph{Importance}: Having an well-defined fee structure will motivate the intermediary/intermediaries to participate in virtual channel creation.  
3) \textbf{Privacy}: In any virtual channel construction, the real identity of a node should only be known to its immediate neighbor(s) \cite{dziembowski2019perun}. \emph{Importance}: Privacy is important since it helps in preserving topology privacy.
4) \textbf{Offchain dispute resolution}: This metric determines if the transaction disputes in a virtual channel require a blockchain write. 
\emph{Importance}: Ideally, we would want a virtual channel construction to have off-chain dispute resolution since blockchain writes are expensive and time-consuming. 
5) \textbf{Support for multihop virtual channels with several intermediaries}:
A virtual channel is said to be multihop
if it can facilitate payments between a sender and a receiver across a path comprising of several intermediate nodes by constructing a single virtual channel from the sender to the receiver, without establishing virtual channels between any pair of intermediate nodes along the path.
\emph{Importance}: 
This property is ideal since it facilitates payments between any pair of nodes in the network by establishing only a single virtual channel, as opposed to a recursive virtual channel in which a sender/receiver, and the intermediate nodes lock coins in multiple virtual channels at the same time. 

The most significant developments in virtual channels over the years are that newer protocols have incorporated a fees to be paid for the intermediary(ies) that lock coins in virtual channels and virtual channels now offer support for both TC based and UTXO based blockchains. Despite these developments, efficient virtual channel protocol design has three overarching research problems, which are discussed in Section \ref{sec:gaps}.

\begin{figure*}[h!]
    \centering
    \begin{subfigure}[t]{0.45\textwidth}
        \centering
        \includegraphics[width=\textwidth]{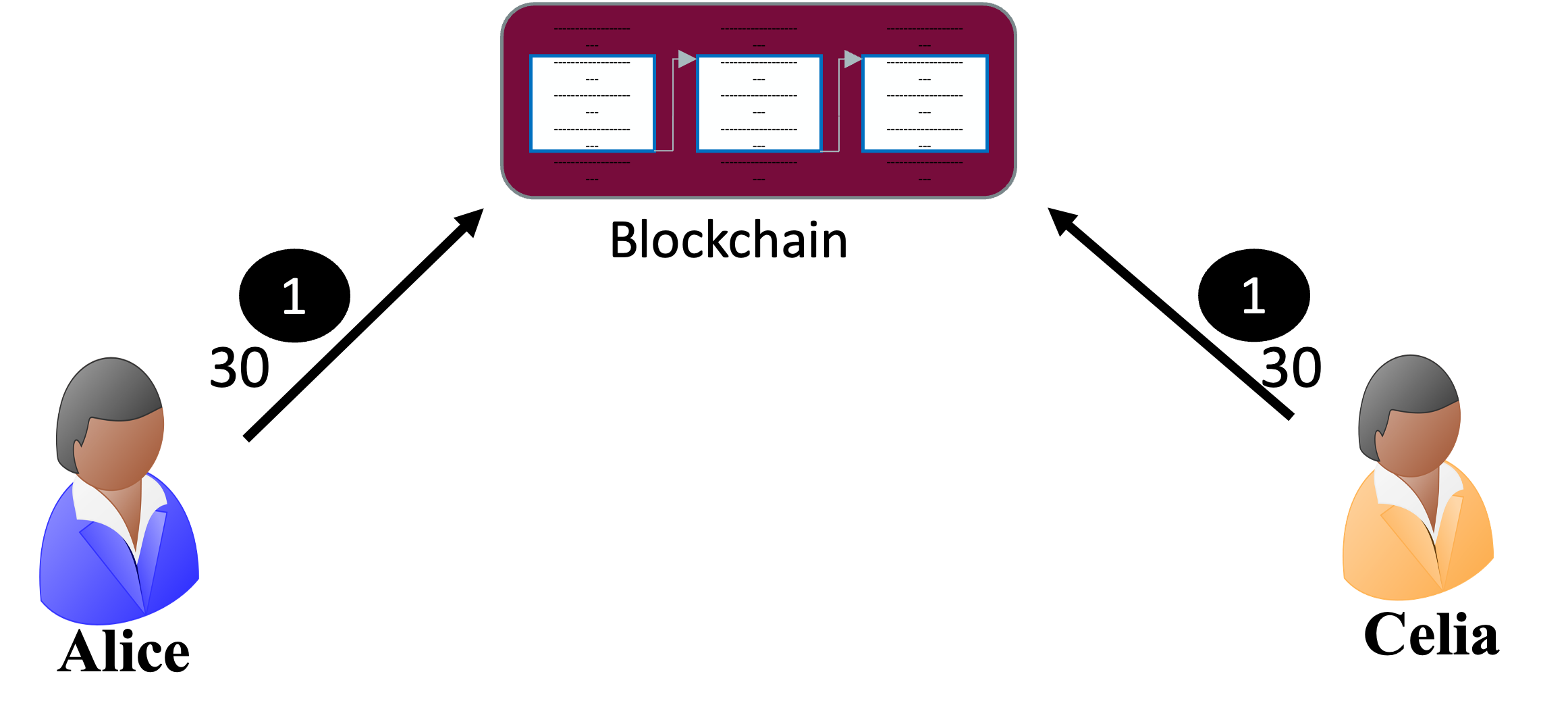}
        \caption{State channel opening}
        \label{fig:statechan1}
    \end{subfigure}
    \hfill
        \begin{subfigure}[t]{0.45\textwidth}
        \centering
        \includegraphics[width=\textwidth]{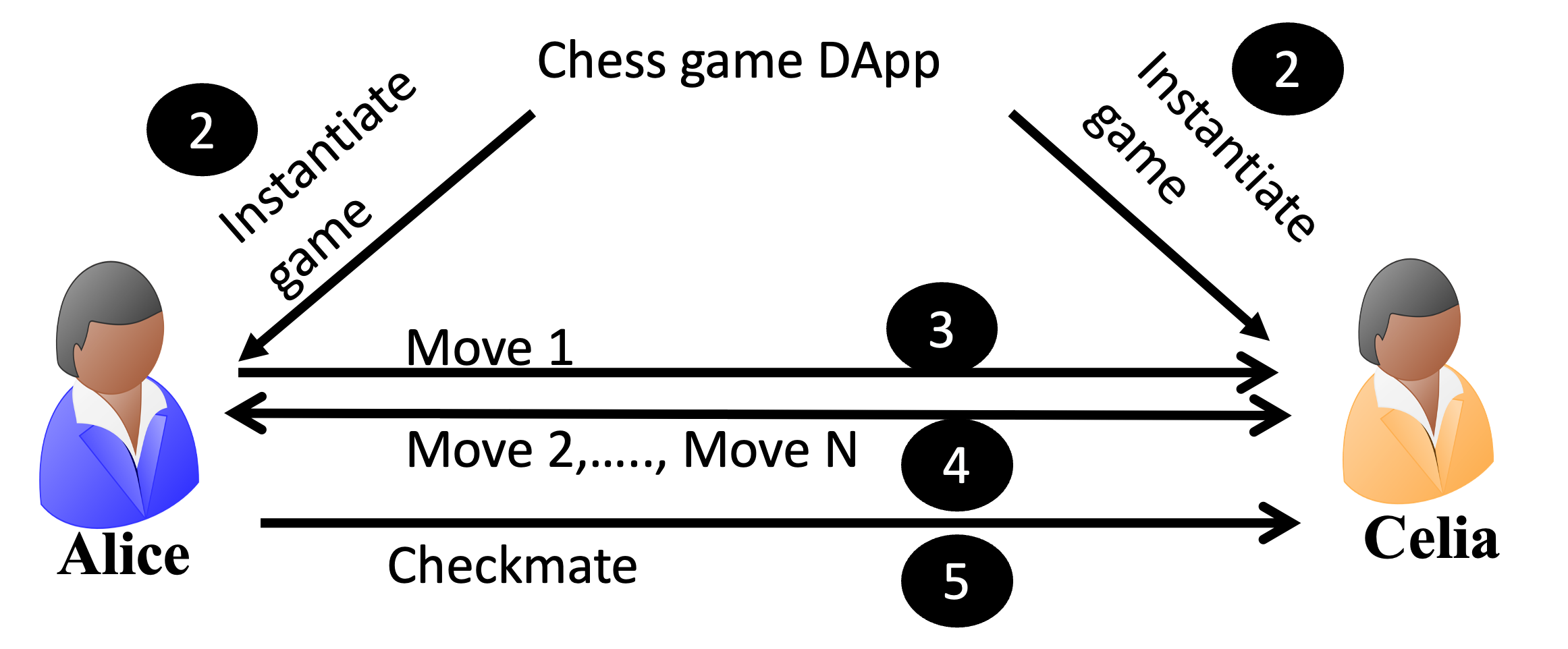}
        \caption{Chess game using state channel}
        \label{fig:statechan2}
         \end{subfigure}
             \hfill
        \begin{subfigure}[t]{0.45\textwidth}
        \centering
        \includegraphics[width=\textwidth]{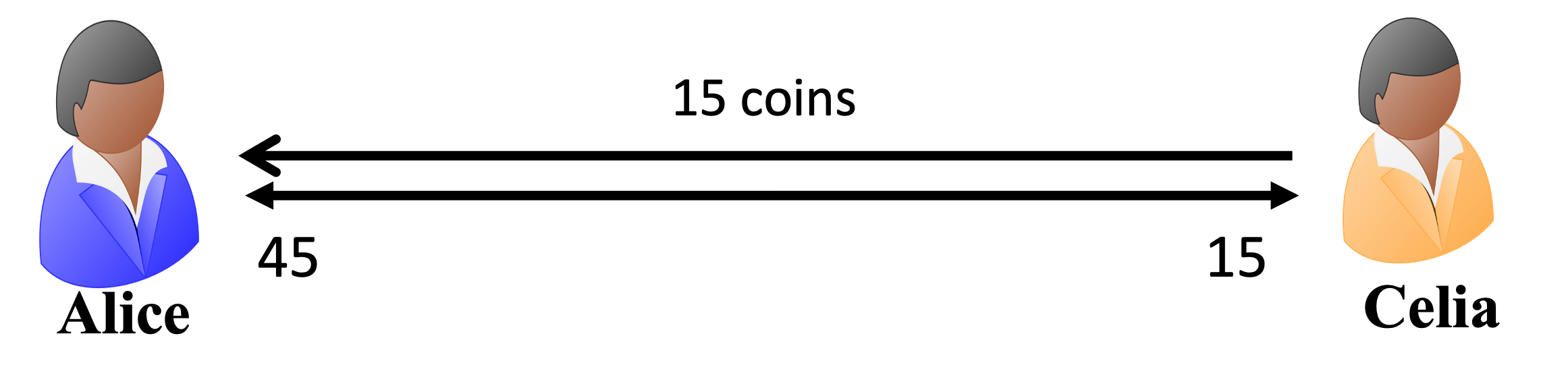}
        \caption{State channel close}
        \label{fig:statechan3}
         \end{subfigure}
         \caption{State channel example with a chess game decentralized application (DApp)}

    \label{fig:combined}
\end{figure*}
\begin{table*}[h!]
\centering
\caption{Comparison of State Channel Protocols}
\label{tbl:statechannel}
\begin{tabular}{|p{1.25cm}|p{1.25cm}|p{1.75cm}|p{1.00cm}|p{1.25cm}|p{1.25cm}|p{1.00cm}|p{1.00cm}}
\hline
Protocol  & BC & \begin{tabular}[c]{@{}l@{}}General/purpose \\ state channels\end{tabular} & Security    & \begin{tabular}[c]{@{}l@{}}Graceful\\ exit\end{tabular}  & \begin{tabular}[c]{@{}l@{}}Off-chain\\ dispute\\ resolution\end{tabular} & Year \\ \hline
\hline
Forcemove \cite{close2018forcemove} & TC         & $\xmark$                                                           & $\checkmark$  & $\xmark$                                                                                                                   & $\xmark$  & 2018                                                               \\ \hline
Pisa \cite{mccorry2019pisa}       & TC   & $\xmark$                                                           & $\checkmark$  & $\xmark$                                                                                                                   & $\xmark$ & 2019                                                                \\ \hline
Sprites \cite{miller2019sprites}   & UTXO       & $\checkmark$                                                       & $\checkmark$  & $\xmark$                                                                                                             & $\xmark$  & 2019                                                               \\ \hline
Hydra \cite{chakravarty2021fast}     & UTXO       & $\checkmark$                                                       & $\checkmark$  & $\xmark$                                                                                                               & $\xmark$   & 2021                                                              \\ \hline

Aumayr \emph{et al.} \cite{aumayr2021generalized}    & UTXO, TC   & $\xmark$                                                           & $\checkmark$  & $\xmark$                                                                                                              & $\xmark$  &2021                                                               \\ \hline
Origami \cite{negka2023origami}   & TC         & $\xmark$                                                           & $\checkmark$  & $\checkmark$                                                                                                               & $\xmark$  & 2023                                                               \\ \hline

\end{tabular}
\end{table*}
\section{State channels}
\label{sec:state}
A state channel is an off-chain protocol that can facilitate execution of an arbitrary decentralized application, also called as DApp (such as a two player game) between two users in a decentralized and distributed network. 
An example of a state channel is given in Figure \ref{fig:statechan1}. Let us consider two users Alice and Celia who intend to play a game of chess. They initially interact with the chess game application deployed on the blockchain by a third party service provider. Both Alice and Celia deposit 30 coins each from their possession into a 2-2 multi signature transaction (step 1). Upon locking the funds, the chess game is instantiated between Alice and Celia by the DApp (step 2). Alice and Celia engage in a series of moves for the chess game (step 3, step 4) and let us assume Alice eventually wins (step 5) as shown  in Figure \ref{fig:statechan2}. Alice gets paid 15 coins as the prize money in Figure \ref{fig:statechan3}. In this game, Alice and Celia countersign each other's moves until there is a winner or a draw. In the event of malicious activity by one party (such as undoing a prior move), the honest party will broadcast all the moves to the blockchain for dispute resolution.\\
\textbf{Motivation}: State channels allow the already existing payment channels to facilitate execution of arbitrary applications such as games, e-commerce, etc. This leads to the PCN becoming more versatile.
State channels should be deployable on any blockchain regardless of the underlying programming/scripting requirements and 
should be able to facilitate the execution of any decentralized application between the parties involved, while maintaining comparable security properties (to payment channels) and should ideally be able to resolve disputes without accessing the blockchain. We now give the properties desired from an ideal state channel, and compare the works in this area on the extent to which they achieve these properties.
We give a qualitative comparison of various state channel protocols in Table \ref{tbl:statechannel}. We describe the metrics used for comparison below. \\
\textbf{Ideal properties}: 1) \textbf{General purpose state channels}: This is the ability of the state channel to facilitate the execution of any application 
supported by the underlying blockchain in an off-chain manner \emph{Importance}: This is an ideal feature since it facilitates the deployment of any application without accessing the blockchain, making state channels more versatile. 
\noindent
2) \textbf{Graceful exit}: A state channel protocol is said to employ a graceful exit if it has clear and well-defined mechanisms for nodes joining or leaving the state channel without accessing the blockchain. \emph{Importance}: This is an important property since most of the state channel protocols require the user to interact with the blockchain when they join/leave the state channel(s) they are a part of.
3) \textbf{Off-chain dispute resolution}: This property indicates if a dispute that occurs between the parties involved in the state channel can be resolved without accessing the blockchain. \emph{Importance}: Ideally, we would want a state channel construction to have off-chain dispute resolution since blockchain writes are expensive and time-consuming. \\
\section{Tumblers}
\label{sec:tumbler}
\begin{table*}[h!]
\centering
\caption{Comparison of Payment Channel Hubs}
\label{tbl:tumble}
\begin{tabular}{|c|c|c|c|c|c|c|c|c|c|}
\hline
Protocol & BC                                             & \begin{tabular}[c]{@{}c@{}}Relationship\\ anonymity\end{tabular} & \begin{tabular}[c]{@{}c@{}}Privacy against\\ aborts\end{tabular} & \begin{tabular}[c]{@{}c@{}}Independent of\\ epochs\end{tabular}  & \begin{tabular}[c]{@{}c@{}}Dynamic\\ corruption\end{tabular} & Atomicity    & \begin{tabular}[c]{@{}c@{}}Value \\ privacy\end{tabular} & \begin{tabular}[c]{@{}c@{}}Variable \\ amount\end{tabular} & Year \\ \hline
\hline
BOLT \cite{green2017bolt}     & TC                                                     & $\checkmark$                                                     & $\xmark$                                                         & $\checkmark$     & $\xmark$                                                     & $\checkmark$ & $\checkmark$                                             & $\checkmark$ & 2017                                              \\ \hline
Tumblebit \cite{heilman2017tumblebit} & UTXO & $\checkmark$ & $\xmark$ & $\xmark$ & $\xmark$ & $\checkmark$ & $\xmark$ & $\xmark$ & 2017 \\
\hline
Nocust \cite{khalil2018commit}   & TC                                                     & $\checkmark$                                                     & $\xmark$                                                         & $\xmark$ & $\xmark$                                                     & $\checkmark$ & $\xmark$                                                 & $\checkmark$ & 2018                                              \\ \hline
Teechain \cite{teechain} & TC                                                     & $\checkmark$                                                     & $\xmark$                                                         & $\xmark$     & $\xmark$                                                     & $\checkmark$ & $\checkmark$                                             & $\checkmark$ & 2019                                              \\ \hline
$A^{2}L$ \cite{a2l}      & UTXO, TC & $\checkmark$                                                     & $\xmark$                                                         & $\xmark$ & $\xmark$                                                     & $\checkmark$ & $\checkmark$                                             & $\xmark$  & 2021                                                 \\ \hline
Accio \cite{accio}    & TC                                                     & $\checkmark$                                                     & $\checkmark$                                                     & $\checkmark$ & $\xmark$                                                     & $\checkmark$ & $\checkmark$                                             & $\checkmark$  & 2021                                             \\ \hline
Boros \cite{zhang2021boros}    & TC                                                     & $\xmark$                                                         & $\xmark$                                                         & $\checkmark$ & $\xmark$                                                     & $\checkmark$ & $\xmark$                                                 & $\checkmark$  & 2021                                             \\ \hline

MIXCT \cite{du2022mixct}    & TC                                                     & $\checkmark$                                                     & $\xmark$                                                         & $\xmark$ & $\xmark$                                                     & $\checkmark$ & $\checkmark$                                             & $\checkmark$ & 2022                                              \\ \hline
Turbo \cite{he2022turbo}    & TC                                                     & $\checkmark$                                                     & $\xmark$                                                         & $\xmark$ & $\xmark$                                                     & $\xmark$     & $\xmark$                                                 & $\xmark$      & 2022                                             \\ \hline

Blindhub \cite{blindhub} & UTXO, TC & $\checkmark$                                                     & $\xmark$                                                         & $\xmark$ & $\xmark$                                                     & $\checkmark$ & $\checkmark$                                             & $\checkmark$   & 2023                                            \\ \hline
\end{tabular}
\end{table*}
\textbf{Motivation}: A payment channel hub (tumbler) is a multi-party off-chain system where users can establish payment channels with a central hub, which acts as an intermediary. It allows multiple users to send payments to each other without the need for direct payment channels between each user pair. The hub coordinates payments between different participants.
The intermediary  which facilitates payments is called a tumbler. 
Though a payment channel hub uses the same underlying infrastructure as that of PCNs and virtual channels, each of these constructions have their own use cases. PCNs are usually used when two nodes Alice and Bob transact on an infrequent basis.
Virtual channels are used if Alice and Bob 
transact frequently, e.g., if Bob provides Alice with a service every month. Payment channel hubs are used when Alice needs to pay several receivers on a frequent basis and she does not want the tumbler to know the receivers.

Payment channel hubs can be classified into two types: onchain and off-chain. Early hubs, also called hubs or mixers, were on-chain \cite{onchain1,onchain2,onchain3,onchain4,onchain5,onchain6,onchain7,onchain8,onchain9,onchain10,meiklejohn2013fistful,maxwell2013coinswap}, but they all suffered from scalability issues due to having to post each transaction on the blockchain. The scalability issues of on-chain payment channel hubs have led to the development of offchain payment channel hubs for specific blockchains, e.g., Bolt~\cite{green2017bolt} for Zcash, Nocust~\cite{khalil2018commit} and MixCT~\cite{du2022mixct} for Ethereum. The most general-purpose payment channel hubs are Tumblebit~\cite{heilman2017tumblebit}, $A^{2}L$~\cite{a2l}, and Blindhub~\cite{blindhub}.
A payment channel hub should be able to facilitate payments between a sender and receiver, who do not have a payment channel open between them such that the hub cannot link a given transaction amount to a particular sender/receiver pair. Furthermore, the payment channel hub should also guarantee the fundamental property of atomicity (ensuring that the payment is sent to the receiver or it does not go through at all). We now give the properties desired from an ideal payment channel hub, and compare the works in this area on the extent to which they achieve these properties. In Table \ref{tbl:tumble}, we qualitatively compare several off-chain tumblers.

\textbf{Ideal properties}:
1) \textbf{Relationship anonymity}: It ensures that the relationship between a sender and a receiver for a given transaction should not be known to the tumbler \cite{green2017bolt}. 
\emph{Importance}: This is an important property since one of the main uses and design goals of a tumbler is to facilitate anonymous transactions. Hence, any tumbler construction should satisfy this property.
2) \textbf{Privacy against aborts}: Tumbler should not be able to deduce the identities of a sender/receiver in case of a transaction abort, regardless of which party is responsible for the abort \cite{blindhub}. 
\emph{Importance}: This is an important property to have since aborts can happen due to several reasons such as network disconnections, power outages, etc. 
Malicious nodes can also deliberately abort a protocol.
In the case of tumbler protocols, the nodes whose payments did not succeed can be linked to each other once the protocol execution completes. 3) \textbf{Independent of time epochs}: The tumbler processes transactions in discrete fragments of time called epochs, i.e., transactions only take place during a time epoch \cite{green2017bolt}. This is not ideal. 
\emph{Importance}: A payment channel hub that supports transaction processing based on time epochs is not efficient because nodes cannot process payments between time epochs, which can be of arbitrary length. The duration between time epochs depends on several factors such as the unit of the time epoch (absolute time or relative time measured in terms of block height), number of users that will most likely involved in the next time epoch, etc. 
4) \textbf{ Protection against dynamic corruption of nodes}: Current tumbler protocols corrupt nodes at the beginning of protocol execution and assume that the set of corrupted nodes remains constant until the protocol execution terminates.
Ideally, a payment channel hub should be able to handle the deviation of any party from the protocol at any point during its execution. 
\emph{Importance}: This is an important property since its infeasible from a practical standpoint to assume that the set of adversaries in the payment channel hub protocol remains constant during execution.
5) \textbf{Atomicity}: Atomicity is ensured 
if either the  payment is routed all the way from the sender to the receiver or the payment is routed at all. 
\emph{Importance}: This is important since it prevents honest parties from losing their funds because of malicious parties.
6) \textbf{Value privacy}: Value privacy is ensured by a payment channel hub, when, given a transaction amount, the tumbler cannot link it to a sender/receiver pair \cite{green2017bolt}. 
\emph{Importance}: Value privacy is important since only the sender and receiver should know the amount being transacted. In the event of every sender/receiver pair transacting a unique amount, lack of value privacy will enable the tumbler to link a sender/receiver pair to a particular transaction, violating relationship anonymity. 
7) \textbf{Variable amount}: The tumbler should be able to process transactions of any amount \cite{green2017bolt}. \emph{Importance}: A tumbler having the ability to process amounts of variable value is important since its practically infeasible to assume that all sender/receiver pairs will always transact a fixed amount.\\
Over the years, various payment channel hub constructions have been developed to address specific challenges based on their unique design goals. However, we have identified three overarching research problems in payment channel hubs that require attention. We discuss them in Section \ref{sec:gaps}.\\

\begin{table}[H]
\centering
\caption{Comparison of Rebalancing Protocols; C is centralized and D is distributed}
\label{tbl:rebal}
\centering
\begin{tabular}{|p{1.30cm}|p{1.00cm}|p{0.75cm}|p{1.00cm}|p{0.75cm} |p{1.00cm}|p{1.75cm}|p{1.25cm}|p{1.25cm}|p{1.00cm}|}
\hline
Protocol              & BC                 & Trusted entity & Graph compatability        & Privacy & Year  \\ \hline
\hline

Revive \cite{revive} & TC & $\checkmark$ (C) & Cycles only &  $\xmark$ & 2017    \\
\hline 
Subramanian \emph{et al.} \cite{subramanian2019rebalancing} & TC and UTXO & $\xmark$ & Agnostic & $\checkmark$ & 2019 \\ 
\hline
Rebal \cite{awathare2021rebal} & UTXO & $\xmark$ & Cycles only & $\checkmark$ & 2021\\
\hline
Hide \& Seek \cite{avarikioti2022hide} & UTXO &  $\checkmark$ (D) & Cycles only & $\checkmark$  & 2022  \\
\hline

Cycle \cite{cycle} & TC & $\xmark$ & Cycles only & $\checkmark$ & 2022 \\ 
\hline

Shaduf \cite{ge2022shaduf} & TC & $\xmark$ & Agnostic & $\checkmark$ & 2022 \\ 
\hline
Musketeer \cite{avarikioti2023musketeer} & UTXO & $\xmark$ & Cycles & $\xmark$ & 2023 \\ 
\hline
Chen \cite{chengraph} \emph{et al.} & UTXO & $\xmark$ & Cycles & $\xmark$ & 2024 \\ 
\hline

\hline
\end{tabular}
\end{table}
\section{Rebalancing} 
\label{sec:rebal}
\textbf{Motivation}: Rebalancing in PCNs involves redistributing the balance within an existing payment channel or across multiple channels to maintain or improve the network’s liquidity and usability. This is done to ensure that both parties in the channel have sufficient funds on their respective sides to send and receive payments. Rebalancing is crucial to payment channel networks. If the link weights of nodes become zero/depleted as a result of being involved in transactions, it will prevent them from taking part in further payments until those link weights are replenished. Such nodes are called as dormant nodes in the network. A PCN which has a large number of such dormant nodes experiences an overall reduction in transaction throughput and an overall reduction in liquidity. Hence, the link weights of the nodes in the PCN need to be rebalanced to prevent failure of transactions due to lack of liquidity.
A rebalancing protocol should be be able to replenish the link weight of dormant nodes in the PCN irrespective of the PCNs topology, while not having to employ any trusted entity and should not compromise privacy of nodes. We now give the properties desired from an ideal rebalancing protocol, and compare the works in this area on the extent to which they achieve these properties. We give a qualitative comparison of rebalancing protocols in Table \ref{tbl:rebal}. We describe the metrics as follows: \\
\textbf{Ideal properties}: 
1) \textbf{Trusted entity}: This metric determines the nature of trust required among parties to deploy the rebalancing protocol. Centralized trust (C) indicates the presence of a single trusted entity and distributed trust (D) indicates the distribution of trust across various entities. \emph{Importance}: It is ideal to have distributed trust for a rebalancing protocol since, PCN payments are by nature decentralized.  
2) \textbf{Graph compatibility}: This indicates the nature of the topology of the network on which a rebalancing protocol can be deployed. Cycles indicate that the rebalancing protocol can only deployed on cyclic graphs and agnostic indicates that the rebalancing protocol can be deployed irrespective of the network topology. \emph{Importance}: Ideally a rebalancing protocol should be deployable irrespective of the topology of the graph since its practically infeasible to assume that every node in a PCN will be a part of a cycle. 
3) \textbf{Privacy}: Privacy is said to be achieved when sensitive information such as the local balance and the identities of nodes are not known to anyone except for their immediate neighbors. \emph{Importance}: The importance of privacy for layer-2 protocols has already been described in Section \ref{sec:routing}. 

Rebalancing is a relatively mature area and significant challenges have already been addressed.


\section{Research Gaps \& Open Problems}
\label{sec:gaps}
In this section, we highlight the gaps in research published up until now in the areas of pathfinding and routing, virtual channel construction, state channels and payment channel hubs. The gaps are described as research questions, denoted by \textbf{RQ}.\\
 \textbf{RQ1: Why is super node liquidity validation in PCNs hard?} A super node, variously called as a trampoline node, routing node, routing helper, landmark node, router, etc.~\cite{blanc,Malavolta2017SilentWhispersES,spider,panwar2024sprite,kolachala2023raced} is a highly connected node with numerous high liquidity channels,
 that helps in pathfinding and routing payments.
One of the main problems with the super nodes is that a sender has no way to know whether the super node possesses enough liquidity on its channels to route a payment. The local balance of a super node in a given channel (or of any node in a PCN) is a private value and should not be known to any node except for its immediate neighbor that it shares the channel with. Currently, if a super node does not have enough liquidity to route the payment of a sender, the transaction fails and it has to be retried by the sender. In $\lightning$, one of the most widely used PCN, this is a significant problem. 
Sometimes the sender might have to keep retrying for $\approx$ 1 hour to have a successful payment \cite{acinqgithub}. The main goal of PCNs is to facilitate instantaneous payments and these transaction retries render such payments almost impossible. It will greatly benefit the sender if it has a mechanism to validate whether a super node has enough liquidity (balance) to route its payment without violating any privacy concerns.\\
 \textbf{RQ2: Why is channel verification in a PCN hard?}
To be a part of any PCN, nodes will open payment channels with other nodes in the network. Two nodes open a payment channel between them by posting a transaction to the blockchain. This transaction can be posted on the blockchain or as a function call to an existing smart contract. In the most popular PCN, $\lightning$, the procedure of verifying whether a payment channel really exists on the blockchain is very inefficient. A node who wants to verify a channel needs to request the block in which the channel opening transaction has been included, verifying whether the transaction has been successfully executed by the validator/miner and finally verifying if the channel opening transaction corresponds to a 2-2 multi signature address on the blockchain. The verifier performing these steps is inefficient since all these steps will have already been performed by the miner. 
Finding a way to do this without blockchain access and in a blockchain agnostic manner in a hard research challenge.\\
 \textbf{RQ3: Why is designing pathfinding protocols for PCNs, that comprise of several distinct well-connected components a hard problem?} Though solutions such as \cite{kolachala2023raced} exist that solve this problem to a certain extent by using routing helpers/trampoline nodes, the aspect that makes it hard is to quantify the denseness/sparseness of a well-connected component. There may be well-connected components in the PCN that comprise of only a few nodes (i.e, islands). In such a case, the nodes in the islands would ideally need to establish payment channels with either a trampoline in their component or a non-trampoline node belonging to other well-connected components. If there is no trampoline available for a well-connected component, the nodes in that component might have to establish payment channels with ideally more than one node from a well-connected component that has a trampoline. This directly contradicts the advantage of payment channels which is to facilitate payments without accessing the blockchain.

 \textbf{RQ4: Why is designing a routing protocol that supports concurrent payments and is resilient to channel gaming a hard problem?}
    Processing concurrent transactions requires the design of a mechanism that allows a node to lock a portion of its liquidity in a channel with an immediate neighbor for one transaction while simultaneously using the remaining liquidity to process another. Though there are protocols that support concurrency \cite{coinexpress,Malavolta2017SilentWhispersES,blanc,kolachala2023raced,ababneh2024auroch,panwar2024sprite}, they are not resilient to  the presence of potentially malicious nodes in the PCN, which may initiate transactions with the sole intent of locking liquidity, leading to congestion and disruption in the network. 
\noindent
 \textbf{RQ5: Why is having a well-defined fee structure for virtual channels hard?} The intermediary(s) involved in the virtual channel construction additionally lock coins in virtual channels apart from the ones locked in the underlying payment channel. Currently, nodes get paid routing fees for every transaction they process.  In the case of virtual channels, having a well-defined fee structure is difficult due to the following reasons: 1) The fee structure should take into account the amount of funds and the time for which these funds of the intermediary(ies) are locked in a virtual channel. 2) It also needs to take into account the routing fee an intermediary could have earned by not locking up the coins in the virtual channel.\\
\noindent
 \textbf{RQ6: Why is off-chain dispute resolution in virtual channels hard?}
 There is no offchain consensus mechanism for dispute resolution in a PCN, as opposed to the 51\% honest majority assumption that exists among validators on the blockchain. This honest majority helps resolve disputes in the transactions posted to the blockchain. 
Designing such a dispute mechanism for layer-2 transactions is hard since transactions are private (not posted to the blockchain), and nodes do not broadcast their activities to the entire network. \\
 \textbf{RQ7: Why is providing support for a multihop virtual channel a hard problem?} This is hard since a multihop virtual channel construction should ensure that neither sender/receiver nor the intermediate nodes should lock coins in multiple channels at the same time.\\
\noindent
\textbf{RQ8: Why is ensuring privacy in a virtual channel protocol a hard problem?} In a recursive virtual channel, new virtual channels are constructed on top of existing virtual channels to facilitate payments. This staggered nature makes it mandatory to reveal the identity of at least one endpoint node (sender or receiver). This is because, at least one node among the sender/receiver is involved in all virtual channels. 
The solution to this problem is to design a multihop virtual channel.\\
 \textbf{RQ9: Why is designing a payment channel hub (PCH) that is resistant to privacy against aborts and dynamic corruption a hard challenge?}
PCHs usually use transaction mixing for enhancing privacy, which is a process in which multiple payments from different users are mixed together in such a way that it is infeasible for the hub to link the sender and recipient of a specific transaction. This process helps obscure the flow of funds, providing unlinkability.
Designing a payment channel hub that is resistant to privacy against aborts is hard because, if a PCH selectively aborts a payment from a sender/receiver, the counter party whose payment also failed can be linked.
 If sender/receiver gets corrupted during the PCH's execution, the corresponding transaction has to be aborted to ensure atomicity, which is why the existing tumbler constructions assume a static adversary, in which certain nodes are designated as corrupted before the PCH begins execution. The trade off here is preserving transaction unlikability during a corrupted party's transaction abort.
\noindent
 \textbf{RQ10: Why is having an offchain dispute resolution for a state channel hard?} Current state channel protocols use an onchain transaction or a function call to the onchain smart contract in the case of dispute resolution. An ideal state channel protocol should be able to facilitate dispute resolutions in an offchain manner. This is a hard research challenge since disputes on the blockchain are usually resolved by the underlying consensus mechanisms. 
In an offchain scenario, there is no such consensus that guarantees transaction validity. Also, in an offchain protocol such as the state channel, nodes which are not part of the state channel protocol do not know any details of the protocol's execution due to privacy concerns. Onchain transactions on the hand are publicly accessible.
\section{Conclusion}
\label{sec:conc}
In this paper, we qualitatively compared the recent work in various foundational areas of PCN research: pathfinding and routing, virtual channels, state channels, payment channel hubs, and rebalancing protocols. We also discussed the gaps in research in these areas along with reasons why fulfilling those gaps is non-trivial. We hope that this paper motivates researchers to build robust protocols that address these gaps that would go a long way towards building out and developing a decentralized financial ecosystem.
\section*{Acknowledgement}
This material is based upon work supported by the National Science
Foundation under Award No. 2148358, 1914635, 2417062, and the Department of Energy under Award No. DE-SC0023392. Any opinions, findings and conclusions or recommendations
expressed in this material are those of the authors and do not necessarily
reflect the views of the National Science Foundation and the Department of
Energy.

\bibliographystyle{IEEEtran}
\bibliography{references}
\end{document}